\documentclass[a4paper,10pt,twoside]{cpc-hepnp}

\usepackage{multicol}
\usepackage{graphicx}
\usepackage{booktabs}
\usepackage{amssymb,bm,mathrsfs,bbm,amscd}
\usepackage[tbtags]{amsmath}
\usepackage{lastpage}

\begin{document}
%\begin{CJK*}{GBK}{song}

\fancyhead[c]{\small Chinese Physics C~~~Vol. 37, No. 1 (2013)
010201} \fancyfoot[C]{\small 010201-\thepage}

\footnotetext[0]{Received 14 March 2009}

\title{Monte Carlo Simulation of RPC-based PET with GEANT4 \thanks{partly Supported by National Natural Science Foundation of
China (U1232206) }}

\author{%
      ZHOU Weizheng$^{1}\footnote{graduated}$
\quad SHAO Ming$^{1,2;1)}$\email{swing@ustc.edu.cn}%
\quad LI Cheng$^{1,2}$ \quad CHEN Hongfang$^{1,2}$ \quad SUN
Yongjie$^{1,2}$ \quad CHEN Tianxiang$^{1}$ }
\maketitle

\address{%
$^1$ Department of Modern Physics, University of Science and Technology of China, Hefei, 230026, Anhui, China\\
$^2$ State Key Laboratory of Particle Detection and Electronics(IHEP \& USTC),USTC, Hefei, 230026, Anhui, China
}

\begin{abstract}
The Resistive Plate Chambers (RPC) are low-cost charged-particle
detectors with good timing resolution and potentially good spatial
resolution. Using RPC as gamma detector provides an opportunity
for application in positron emission tomography (PET). In this
work, we use GEANT4 simulation package to study various methods
improving the detection efficiency of a realistic RPC-based PET model for
511keV photons, by adding more detection units, changing the
thickness of each layer, choosing different converters and using
multi-gaps RPC (MRPC) technique. Proper balance among these factors are
discussed. It's found that although RPC with materials of high
atomic number can reach a higher efficiency, they may contribute
to a poor spatial resolution and higher background level.
\end{abstract}



\footnotetext[0]{\hspace*{-3mm}\raisebox{0.3ex}{$\scriptstyle\copyright$}2013
Chinese Physical Society and the Institute of High Energy Physics
of the Chinese Academy of Sciences and the Institute
of Modern Physics of the Chinese Academy of Sciences and IOP Publishing Ltd}%

%\begin{multicols}{2}

\section{Introduction}

Positron emission tomography (PET) is a nuclear medicine imaging
technique that provides three-dimensional (3-D) images of a tissue
\cite{pet1}. Short-life radionuclides, such as $^{11}$C
and $^{13}$N, can emit $e^{+}$ into the tissue. After a short
distance, the emitted positron annihilates with an electron. As a
consequence, a pair of photons with about 511keV energy are
produced and move in approximately opposite directions. By
detecting the photons, a 3-D spatial distribution of the
radionuclides can be reconstructed.

One application of PET is on small animal tomography. Considering
their small size, high spatial resolution imaging detectors are
required. However, a traditional PET system is usually based on a
crystal scintillator, which has a limited spatial resolution and
is rather expensive \cite{pet2}. A few years ago, a novel PET
based on Resistive Plate Chambers (RPC) was proposed
\cite{rpcpet1} due to its good timing resolution, potentially good
spatial resolution and low cost.

RPC is a gas detector that has been widely used in high energy
physics for decades \cite{rpc1}. It exploits gas amplification in
a uniform electric field between two resistive parallel plates.
When a charged particle passes through the chambers, gas molecules
in the gap are ionized into electrons and ions. In a strong
electric field, the liberated electron causes an avalanche
(usually with spatial development of sub-mm size). By detecting
the avalanche signal, the position and time when the charged
particle is hit can be recorded. RPC is typically used to detect a
charged particle. Photons, as neutral particles, can be detected
by RPC only after they have been converted to electrons. However,
the efficiency of RPC for 511keV photons is low ($< 1\%$).
Therefore, a major topic on the RPC-PET technique is to enhance its
detection efficiency for 511keV photons.

A small-animal PET system with timing RPC technology has been
built and tested \cite{rpcpet2}. Detailed simulation of
converter-plate stacks made of three materials has been performed,
and efficiency measurements has confirmed the reliability of the
simulation results \cite{rpcpet3}. In Ref. \cite{rpcpet4}, three
designs of RPC have been discussed, the best one is a
sandwich-type gas-insulator-converter detector, with bismuth or
lead converters. It can reach a high efficiency and significantly
suppress the RPC's respond of low-energy photons.

In this paper, we want to further study the performance of a
RPC-PET structure by using Monte Carlo simulation, based on more
realistic detector structure compared to previous works. Our
experience on multi-gap RPC (MRPC)
\cite{mrpc2}\cite{mrpc3}\cite{mrpc4}\cite{mrpc5} is used for this
study. We essentially introduced four methods to improve the RPC
detection efficiency for 511keV photons: adding more detection
units, changing the thickness of each layer, choosing different
converters and using multi-gaps RPC. Furthermore, we found that a
compromise between the detection efficiency and spatial resolution
should be made. At last, several image reconstructions are
performed to check the reconstruction ability of RPC with different converters.

\section{Detector Configureation}

\begin{figure}[ht]
\begin{center}
{\includegraphics[width=0.48\textwidth] {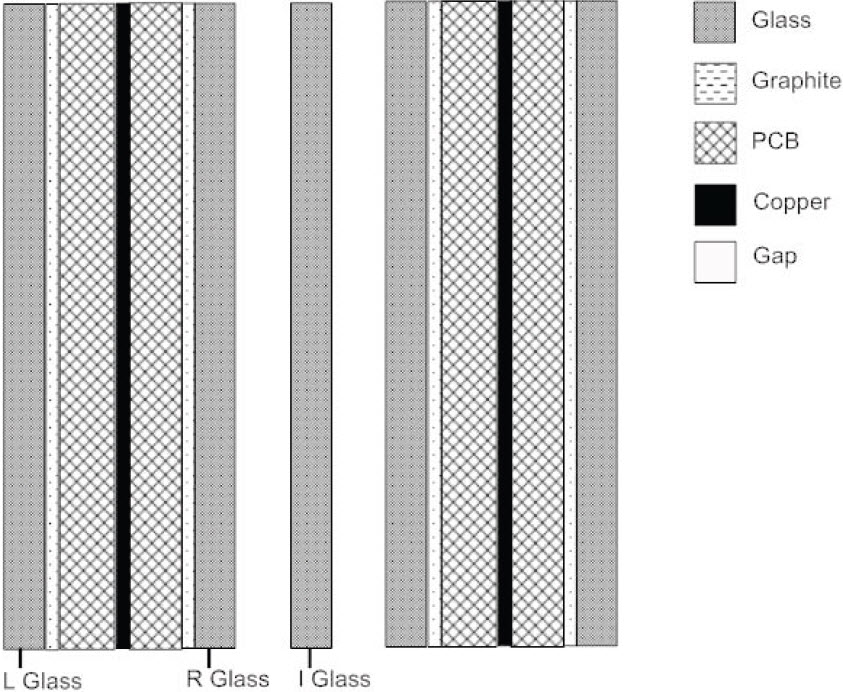}}
\end{center}
\caption[]{Detector structure of the RPC-based PET unit.}
 \label{detUnit}
\end{figure}

The detector configuration is shown in Fig. \ref{detUnit}. A
detection unit, sandwiched between left glass (L Glass) and right
glass (R Glass) plates, is composed of 2 printed circuit boards (PCB),
2 graphite layers and 1 copper (readout pad) layer. Between two
readout layers, one or more inner glass (I Glass) plates can be
inserted. The glasses serve as the highly resistive material,
while graphite layers between the PCB and glass are used as HV
electrode. The gas gap is filled with a mixture of $85\%C_{2}H_{2}F_{4} +
10\%SF_{6} + 5\%C_{4}H_{10}$. The size of RPC is $40 mm \times 40 mm$. The
thickness of each layers can affect the detection efficiency, and will
be discussed later in the article. The gamma source is simplified
as a point-like source emitting 511keV photons, which is placed in
front of the RPC and the direction of incident is set
perpendicular to the surface. The criterion of whether a photon is
detected is that at least one electron converted from the photon,
emerges in the gas gap and deposits non-zero energy. The GEANT4
package (version Geant4-09-02-patch-01) has been used for the
simulation \cite{geant}.
%% In this paper, each run is composed of 400,000 events, wherein one photon is emitted in every event.

\section{Results and discussion}

A \emph{primary} 511keV photon is converted to electron mainly through two processes:
photoelectric effect and Compton scattering. For clarity, various converted
electrons that come directly from the \emph{primary} photon, either through photoelectric effect or
Compton scattering, are called A-type electrons. All other electrons that
come from multiple Compton scattering, with or without photoelectric effect,
are called B-type electrons. The spatial resolution of the former electron
is found to be much better than the latter since the \emph{secondary}
photon deviates from the original direction after undergoing Compton scattering.
When B-type electrons (converted from the \emph{secondary} photons) are detected, we can not
accurately distinguish the original photon direction. Hence, the spatial
resolution deteriorates and the photon pair reconstruction of the PET is
unprecise.

According to the simulation, only $6\%$ of the produced electrons pass
through the gas gap with zero energy deposited. That means if an
electron emerges in the gap, it would cause an avalanche at $94\%$
probability. Meanwhile, a positron can be located by detecting the
pair of photons coming from the annihilation. If the detection
efficiency of RPC for a 511keV photon is $p$ and the
percentage of A-type electrons (among all converted electrons) is $q$,
then $0.94^{2}p^{2}q^{2}$ of all emitted photons are useful for image
reconstruction. It's thus clear that the efficiency and the percentage of A-type
electrons are both important for the PET image reconstruction.

\begin{figure}[ht]
\begin{center}
{\includegraphics[width=0.48\textwidth] {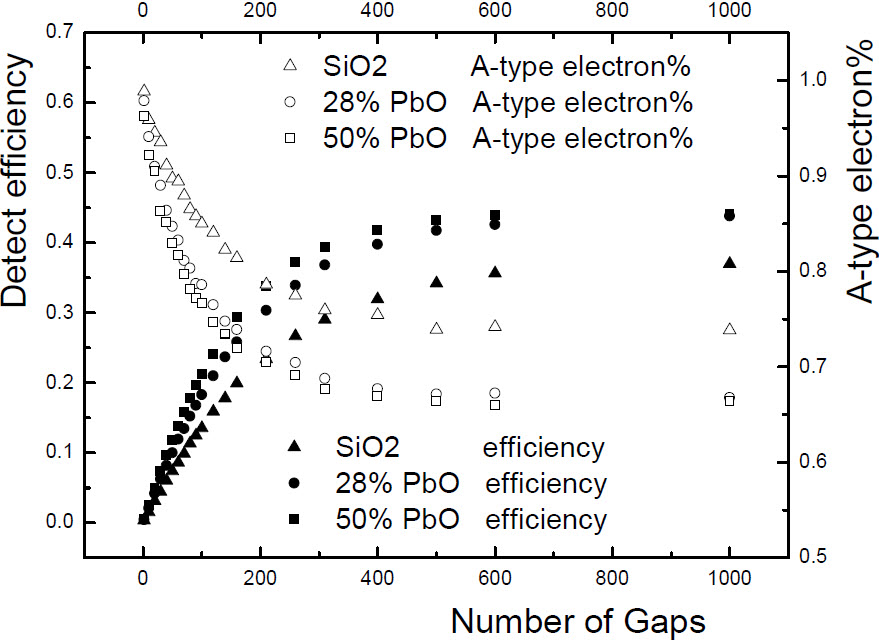}}
\end{center}
\caption[]{Detection efficiency and A-type electrons percentage,
depending on the number of gas gaps.}
 \label{ngap}
\end{figure}

\subsection{Efficiency}

Since the detection efficiency of RPC currently used in high
energy physics experiments for 511keV photons is low, our main
goal is to improve it. We have attempted to vary the parameters
used in our simulation and have found that efficiency mainly
depends on four factors: the number of gaps, the thickness of
material, the thickness of converter, and the number of inner
glasses.

\begin{figure}[ht]
\begin{center}
{\includegraphics[width=0.48\textwidth] {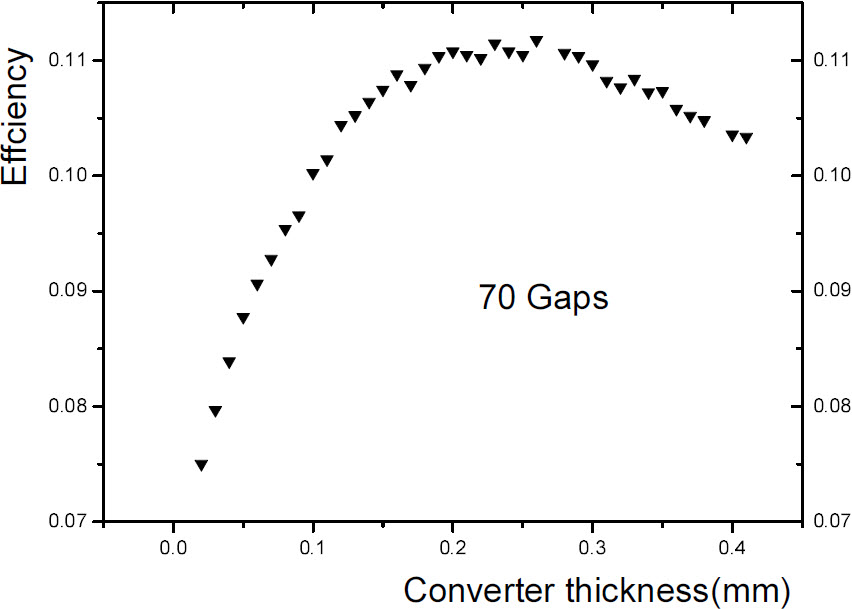}}
\end{center}
\caption[]{Detection efficiency as a function of convertor
thickness.}
 \label{cthick}
\end{figure}

\subsubsection{Number of gaps}

We can enhance the detection efficiency by using more gas gaps, as
shown in Fig. \ref{ngap}. Here, the gaps means the total gaps of all the
RPC units (e.g. the total number of gaps for 45 RPC units with one inner
glass is 90). The efficiency increases with the number of gaps and
reaches saturation afterwards. If the RPC has a large number of
gaps, almost all the photons will be absorbed by RPC so that adding
more units would be meaningless. Also, we can find that RPC with
lead glass has a higher detection efficiency. In Fig. \ref{ngap}, "SiO2"
refers to RPC with normal glass as converter, while $28\%$/$50\%$ PbO means
lead glass convertor containing $28\%$/$50\%$ PbO in weight. In the following
sections we'll choose 90 gaps for further disucssion, considering the feasibility in
producing a real prototype \cite{rpcpet3}\cite{mrpc4}. The configuration of RPC is as
follows: an active area of $40mm\times40mm$, one inner glass,
normal glass as converter, and 90 gaps in total. The thickness of
each layer are: PCB 0.2 mm, Copper 0.035 mm, graphite 0.02 mm, glass
0.1 mm, and gas gap 0.22 mm. In the rest of this paper, this defined
configuration is always used unless otherwise specified.

\begin{figure}[ht]
\begin{center}
{\includegraphics[width=0.48\textwidth] {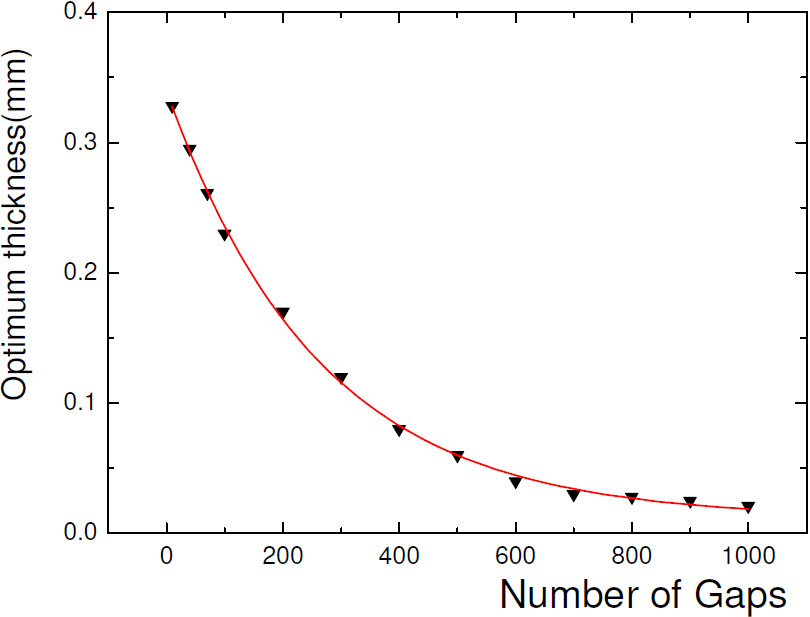}}
\end{center}
\caption[]{Optimum convertor thickness for detection units with
different number of gas gaps. The curve shows an exponential fit.}
 \label{optthick}
\end{figure}

\subsubsection{Thickness}

The dependence of efficiency on the converter thickness was
discussed in Ref. \cite{rpcpet4}. Efficiency increases with converter
thickness until a maximum value is reached, and after which it
will start to decrease. Our simulation results indeed confirms this
conclusion. As Shown in Fig. \ref{cthick}, for a 70-gap RPC the
efficiency reaches its maximum when the thickness of glass is
approximately 0.26mm (optimal thickness), then decreases. Furthermore, we find that the
optimal thickness has an exponential relation to the number of
gaps, see Fig. \ref{optthick}. As the number of total gas gaps increase,
to maximize the detection efficiency the convertor has to be
thinner. A RPC can be simplified as a converter-gap-stacks model,
thus the number of electrons emerging in $n$ gaps reads \cite{rpcpet4}:

$N=\sum_{i=1}^{n}a_{0}(e^{-x/c}-e^{-x/b})e^{-(i-1)x/c}$, with
$a_{0}=N_{0}k/(b-1/c)$.

In this formula, $N_{0}$ is the initial number of photons, $k$
is the photon interaction coefficient, $b$ is the electron interaction
coefficient, and $c$ is the photon beam attenuation coefficient.
All the coefficients $k$, $b$ and $c$ are constants. However, from
this model the exponential relationship in Fig. \ref{optthick} can not be deduced.
Besides, for thin converters, a fraction of electrons
converted in non-converter materials (such as PCB) can
emerge into the gas gap and contribute to the overall efficiency.
This discrepancy may arise from the simplification in \cite{rpcpet4} that
the electron interaction coefficient is constant, which is not true considering eg.
the asymmetric photon energy distribution in forward and backward scatter direction
and the non-linear absorption tail at the end of an electron's range.
Non-converters, such as PCB, copper layer and graphite, can also
influence detection efficiency.Based on an overall consideration
of various factors, the optimal
thickness simply decreases exponentially with the number of gaps. This
simple relationship is useful when designing a practical RPC.

\begin{figure}[ht]
\begin{center}
{\includegraphics[width=0.48\textwidth] {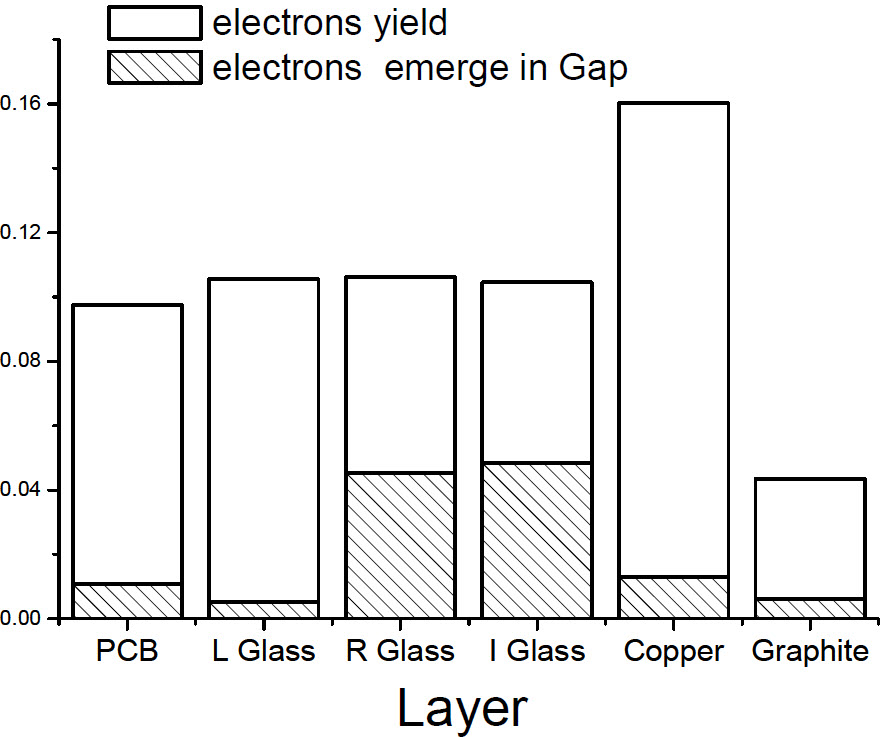}
\includegraphics[width=0.48\textwidth] {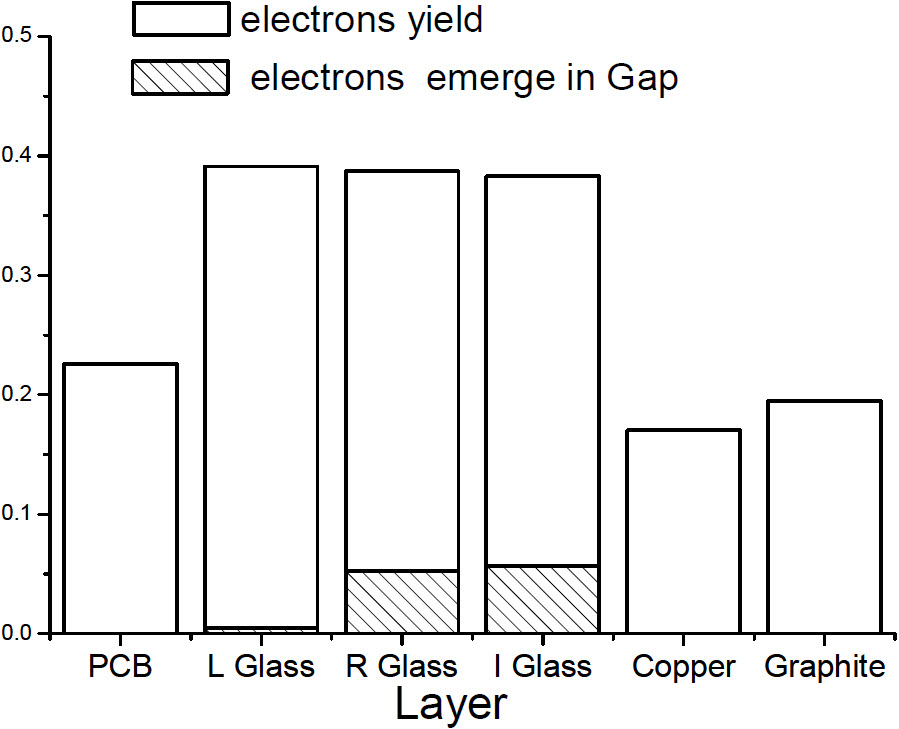}}
\end{center}
\caption[]{Normalized electron yields from all materials in a RPC,
for (a) Configuration: one inner glass, 90 gaps, PCB 0.2 mm,
Copper 0.035 mm, Graphite 0.02 mm, Glass 0.1 mm, Gap 0.22 mm,
using normal glass as converter; and (b) Configuration: one inner
glass, 90 gaps, PCB 0.5 mm, Copper 0.035 mm, Graphite 0.1 mm,
Glass 0.4 mm, Gap 0.22 mm, using normal glass as converter.}
 \label{effCont}
\end{figure}

Detection efficiency depends on
two processes: photon interaction with materials and electron
propagation through materials. Electrons converted in a
non-converter is more likely stopped by glasses and be unable to
emerge in the gap. Thus, these electrons (which are converted
in non-converter materials) make less contribution in photon
detecting, but rather consuming photons. As shown in Fig. \ref{effCont}(a), most of
the electrons converted in a non-converter (PCB, copper, and
graphite electrode) do not emerge in the gap as a result of stopping by
glasses and other materials. In contrast, half of the electrons
converted in inner and right glasses can emerge in the gas gap, thus
be detected. Consequently, the thinner the non-converter
materials are, the higher the efficiency is for a larger number of
gaps. Furthermore, electrons converted in left glass can hardly
emerge in the gap. Therefore, it could be thinner than the I and R
Glasses. In Fig. \ref{effCont}(b) thicker non-converter and converter
are used to compare with thinner materials in Fig. \ref{effCont}(a).
It's found although the efficiency to detect an electron is similar,
the ratio of detected to produced electron is much smaller in Fig. \ref{effCont}(b).
This will affect the final detection efficiency at large gas gap limit.
Note in Fig. \ref{effCont}(b) the number of electrons yield is larger
than the number of original 511keV photons, since one photon can
produce more than one electron. Both the number of electrons yield and
electrons emerging in the gap are normalized by the number of
original 511keV photons in Fig. \ref{effCont}. Hence the sum of electron
yield from all layers can be larger than 1.

\begin{figure}[ht]
\begin{center}
{\includegraphics[width=0.48\textwidth] {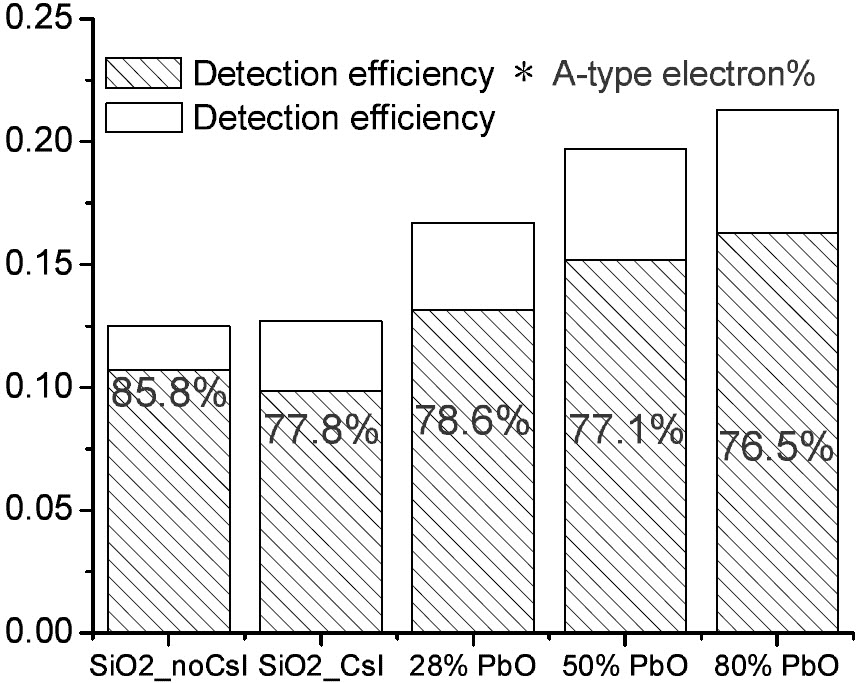}}
\end{center}
\caption[]{The efficiency for RPC with different convertor materials.
The numbers denotes the A-type electron percentage.}
 \label{mConv}
\end{figure}

\begin{figure}[ht]
\begin{center}
{\includegraphics[width=0.48\textwidth] {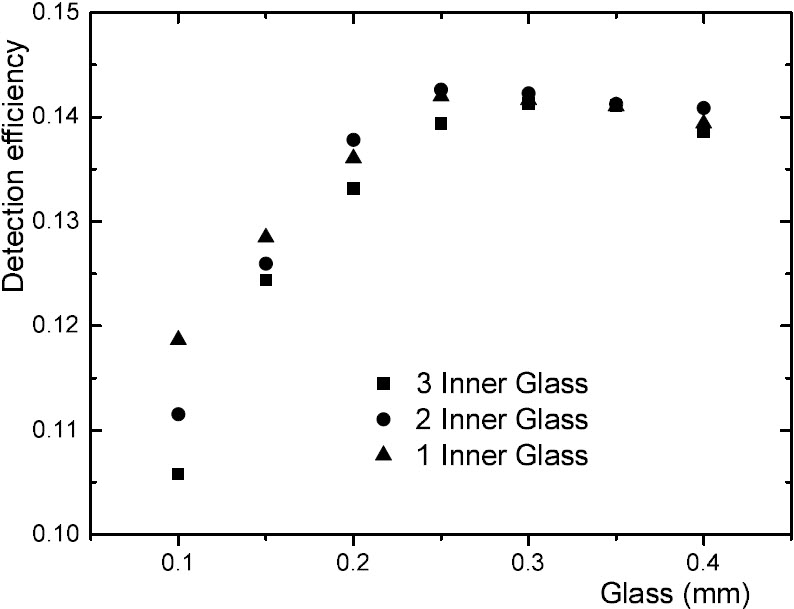}}
\end{center}
\caption[]{The efficiency of MRPC with different numbers of inner
glass. The number of gaps and thickness of each layer (except I glass) are fixed.}
 \label{nIglass}
\end{figure}

\subsubsection{Converter}

The cross-sections for photon interactions with materials increase
with atomic number. Therefore, materials with high atomic number
elements are considered in the simulation. We have attempted to
use several materials as converters: normal glass, lead glass
containing PbO, and normal glass coated with CsI. As shown in Fig.
\ref{mConv}, glass with $10{\mu}m$ CsI coating has only slightly higher efficiency than normal glass,
because the CsI layers not only produce electrons, they also block
electrons from emergin into the gas gaps. For lead glass, although the detection efficiency increase with
the percentage of PbO, the A-type electron percentage decreases (ie. B-type electron percentage increses).
A larger percentage of B-type electrons in
lead glass lead to a poorer spatial resolution. With glass
containing $80\%$ PbO, the overall detection efficiency of RPC reaches
$21.3\%$, while the A-type electron percentage drops to $76.5\%$.

The A-type electron percentage also depends on the number of total gas
gaps, as shown in Fig. \ref{ngap}. Adding more detection units and using
high atomic materials will increase detection efficiency and meanwhile
decrease the A-type electron percentage. At large number of gas gap, the
efficiency and A-type electron percentage curve of $28\%$ PbO
and $50\%$ PbO have only little difference. In such a case it's not sensitive to vary the
percentage (eg. $80\%$) of PbO.

\subsubsection{Number of inner glasses}

Different numbers of inner glass (I Glass) can also influence
efficiency. The efficiency of RPC with different numbers of inner
glass (ie. MRPC) are compared, as shown in Fig. \ref{nIglass}. When the thickness of the
I glass is around the optimal value ($0.25\sim0.3mm$), the RPCs with 1, 2 or 3 inner glasses
show little difference in efficiency.

Note in the simulation the total number of gas gaps and thickness of each layer (except I glass) are fixed,
and normal glass is used as converter.

\begin{figure}[ht]
\begin{center}
{\includegraphics[width=0.48\textwidth] {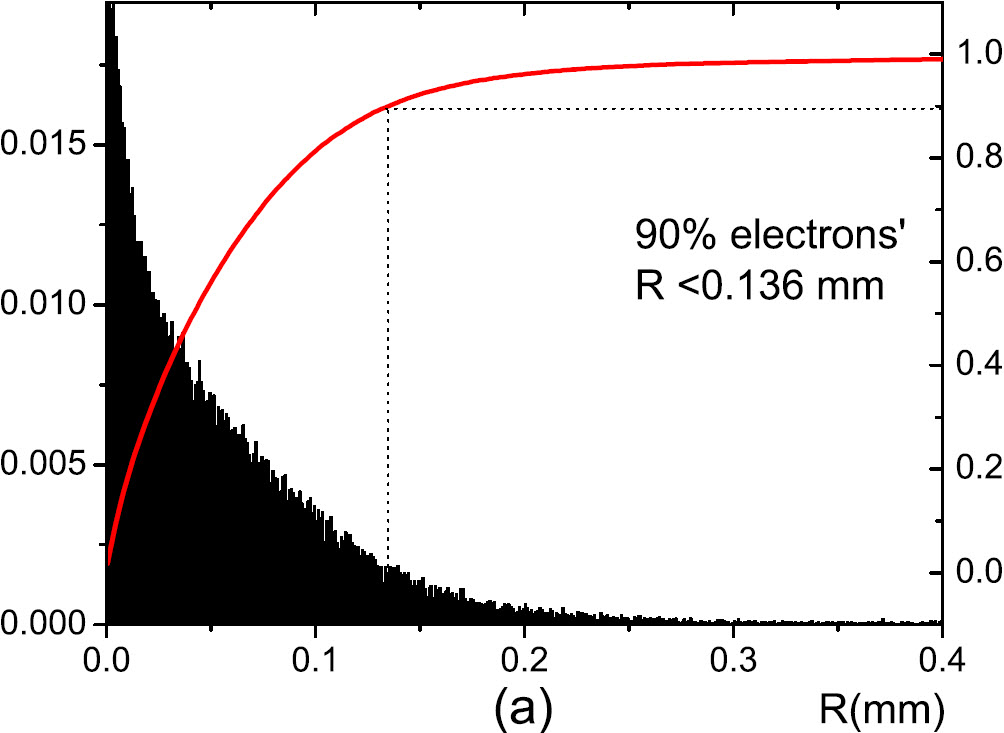}
\includegraphics[width=0.48\textwidth] {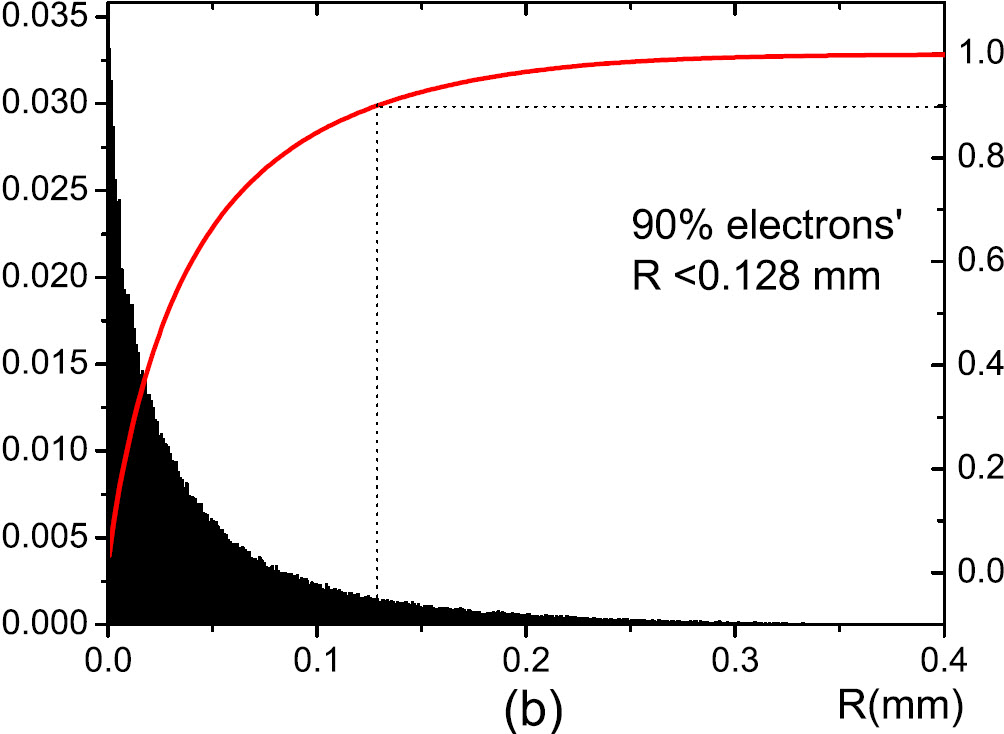}
\includegraphics[width=0.48\textwidth] {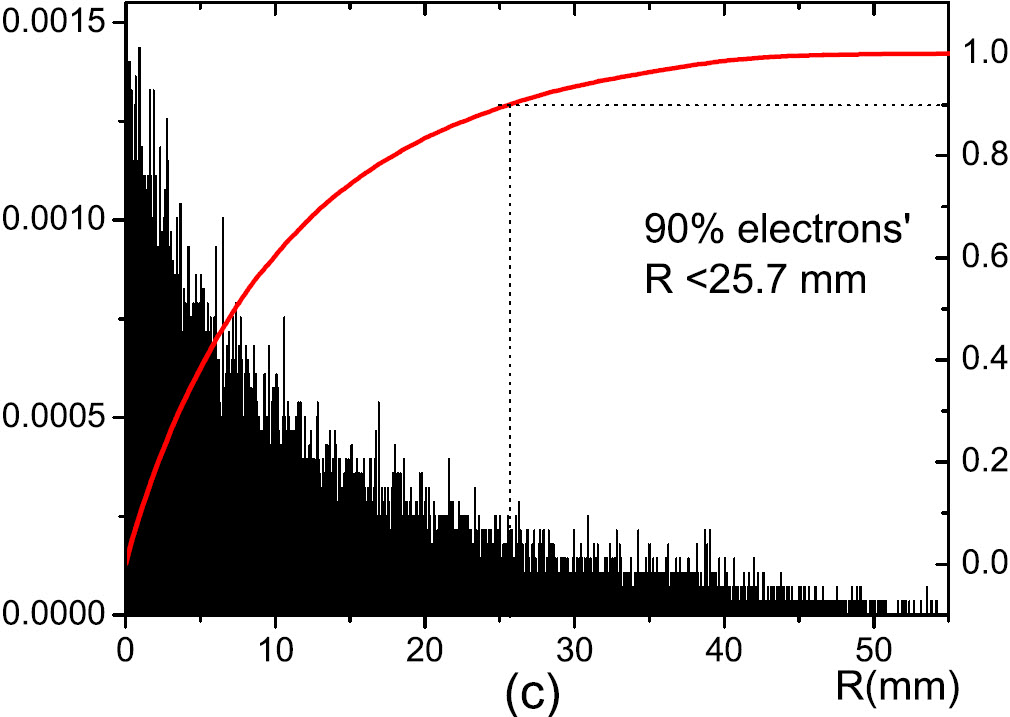}
\includegraphics[width=0.48\textwidth] {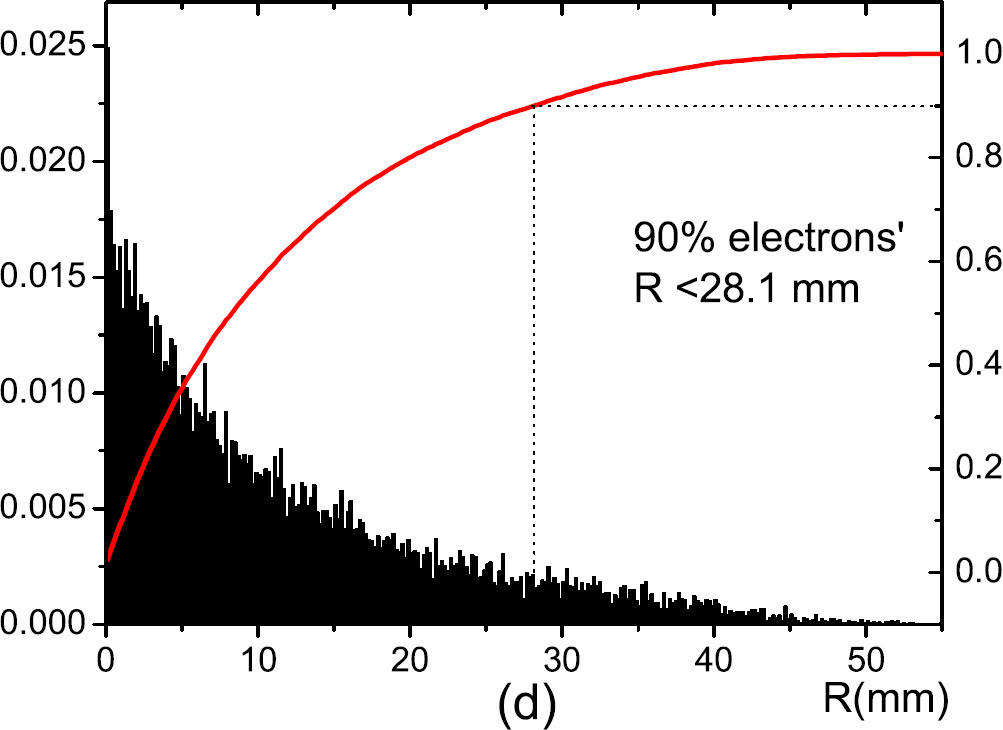}}
\end{center}
\caption[]{The distribution of electron hit position (in gas gap).
a) Electron from photoelectric effect; b) Electron from
a one-time Compton scattering; c) Electron from a two-time Compton
scattering, and the one-time Compton electron is not detected; d)
Electron from photoelectric after a one-time Compton effect, and the
Compton electron is not detected. The curve is the cumulative
histogram of R.}
 \label{pReso}
\end{figure}

\subsection{Spatial resolution}

We have discussed the A-type and B-type electrons previously, and mentioned
the spatial resolution of A-type electrons is much better than
that of B-type electrons. Based on the simulation data, we
calculate the distribution of hit position, the results of which
are as shown in Fig. \ref{pReso}. R is defined as the deviation of the
position when the electron is hit from the incident direction of
the original photon. As shown in Figs. \ref{pReso}(a) and \ref{pReso}(b), the R value
of almost $90\%$ electrons is less than 0.15 mm for the A-type electrons.
However, more than $50\%$ B-type electrons' R value is larger than 10 mm, as shown in
Figs. \ref{pReso}(c) and \ref{pReso}(d). Noticeably, a small percentage of A-type
electrons leads to a much worse spatial resolution. Hence, the
percentage of B-type electrons should be suppressed for good PET reconstruction. A compromise
should be made between the detection efficiency and spatial
resolution. Lead glass may be selected in cases RPC with
fewer gaps are needed (eg. 10 gaps and lead glass of $80\%$ PbO).
If a higher resolution is required, a converter of normal glass
with more gas gaps should be considered (eg. 90 or more gas gaps and
normal glass).

\begin{figure}[ht]
\begin{center}
{\includegraphics[width=0.48\textwidth] {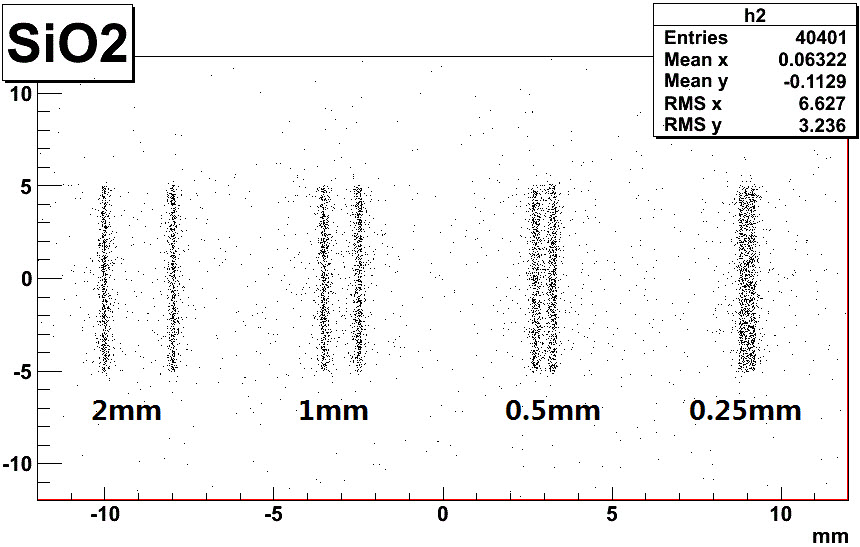}}
\end{center}
\caption[]{Double-line reconstruction.}
 \label{dline}
\end{figure}

\subsection{Image reconstruction}

In order to check the capability of image reconstruction, four
RPCs are placed in the x-y plane with an isotropic 511keV gamma
source at the center of the four RPCs. We are not mainly concerned
with image reconstruction method in this paper. Therefore, we only
use a simple back-projection algorithm for demonstration purpose.
Various sophisticated methods in the field of PET exist, including
the filter back-projection (FBP) algorithm \cite{fbp}, a standard
reconstruction algorithm. Each event is composed of a pair of
511keV photons that are emitted from the center of four RPCs and
move in opposite directions. If both photons are detected, two
hits are recorded. We can then reconstruct the image by connecting
the two hit points.

\begin{figure}[ht]
\begin{center}
{\includegraphics[width=0.48\textwidth] {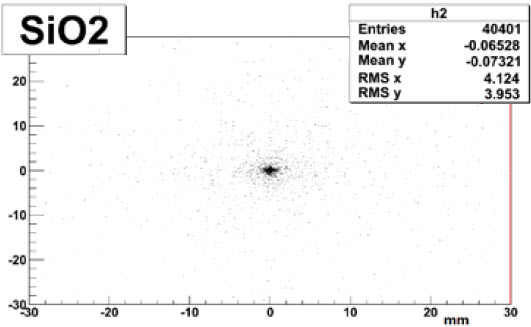}
\includegraphics[width=0.48\textwidth] {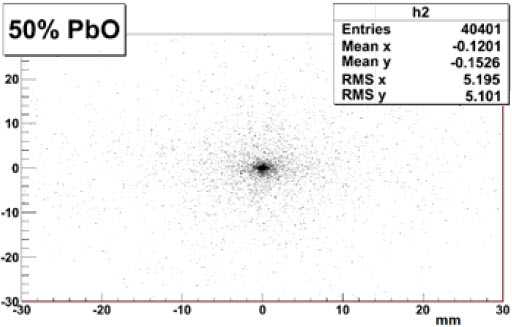}}
\end{center}
\caption[]{Spatial resolution of Reconstructed source
image with different converter material.}
 \label{recReso}
\end{figure}

\begin{figure}[ht]
\begin{center}
{\includegraphics[width=0.48\textwidth] {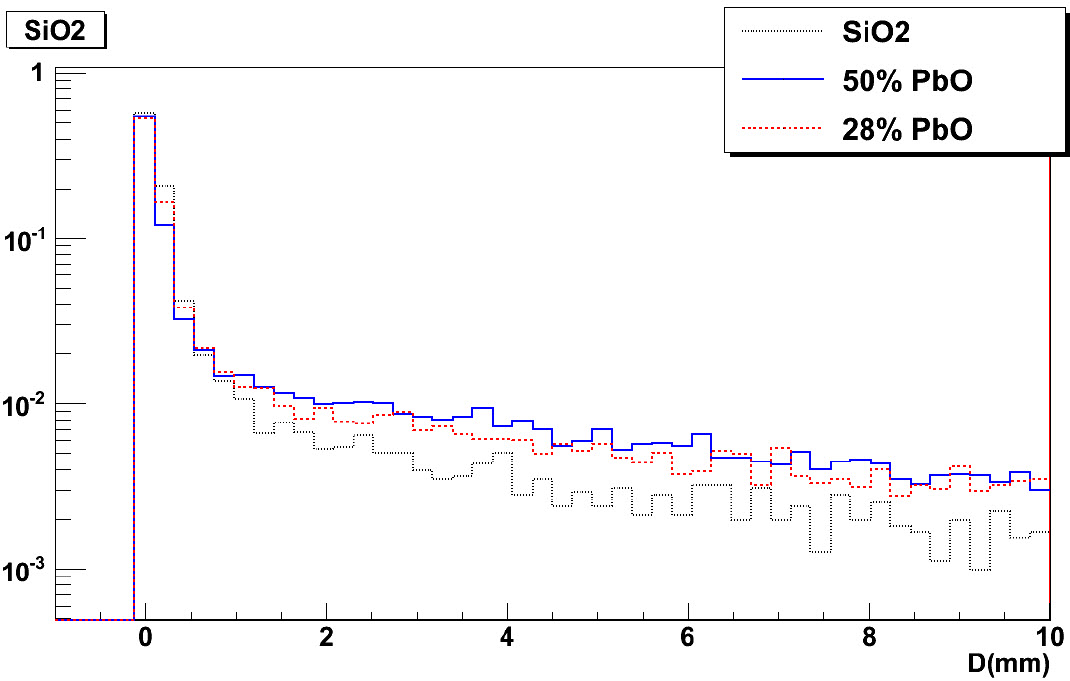}}
\end{center}
\caption[]{Distance between the reconstructed hit point and the original
point.}
 \label{recDist}
\end{figure}

We have performed several reconstructions of double-line sources, as shown in
Fig. \ref{dline}. We can observe that RPC has the ability to distinguish two
lines separated by about $0.5mm$. Please note in Fig. \ref{dline} no other
factors are considered, such as the hit positioning uncertainty of the RPC.

To compare the difference of spatial resolution between lead glass
and normal glass converter, we have performed several reconstructions image
using three different RPC converter materials. The back-projected scatter
points are shown in Fig. \ref{recReso}, while the distributions of D are shown in
Fig. \ref{recDist}. Here, D is defined as the distance between each line
connecting the two hits and the source points, which is at the center of the four RPCs. Lead glass has
definitely a stronger background compared with normal glass. The reconstructed
background level when using $50\%$ PbO converter is approximately $2-3$ times larger
than when using normal glass. So if heavy elements such as Pb, Bi and Au are used in converters, a
strong background may seriously influence image reconstructions. For example, considering that a strong source and
a weak source are close to each other, the high background level
from lead glass by the strong source may cause a low-quality reconstruction of the weak
source. Thus a comprise may be needed
between improving the detection efficiency and decreasing the spatial uncertainty.

\section{Conclusion}

Various methods are attempted to improve the detection efficiency of RPC for 511keV
photons, by adding more gaps, tuning the thickness of the
converter, choosing thin non-converter materials, using converter containing high
atomic number elements, and selecting multi-gaps RPC. Although
RPCs with materials of high atomic number elements can reach a higher
efficiency, they may lead to a stronger background, thus
worsens the spatial resolution. A compromise should then be made
between detection efficiency and spatial resolution. Lead glass
can be chosen as the converter in cases that RPC with fewer gaps
are used and short imaging time is needed. A converter composed of normal glass having more gaps
should be considered if a higher resolution is required. With the
RPC structure described in this paper, when 90 gaps is chosen, the detection
efficiency of RPC with lead glass containing $80\%$ PbO can reach
$21.3\%$, while that of RPC with normal glass is $12.5\%$.
Meanwhile the A-type electron percentage decrease from $85.8\%$ to
$76.5\%$.

\vspace{5mm}

%%\section*{References}

%\end{multicols}

\clearpage

%\end{CJK*}
\end{document}